\documentclass[a4paper,12pt]{article}
\usepackage{amssymb,amsmath,graphics}

\newcommand{\be}{\begin{eqnarray}}
\newcommand{\ee}{\end{eqnarray}}

\newcommand{\fig}[1]{Fig.~\ref{#1}}
\newcommand{\eqn}[1]{Eq.~(\ref{#1})}

\newcommand{\ket}[1]{| #1 \rangle}

\newcommand{\xor}{\oplus}
\newcommand{\pauli}[4]{\left(\begin{array}{cc} #1&#2\\#3&#4
\end{array}\right)}

\begin{document}

\title{ROM-BASED COMPUTATION: \\ QUANTUM VERSUS CLASSICAL }
\author{
\small B. C. Travaglione, M. A. Nielsen \\
\small \it Centre for Quantum Computer Technology, University of Queensland \\ 
\small \it St Lucia, Queensland, Australia \\
\small H. M. Wiseman \\
\small \it Centre for Quantum Dynamics, School of Science, Griffith University
\\
\small \it Nathan, Queensland, Australia \\
\small A. Ambainis \\
\small \it Computer Science Division, University of California \\
\small \it Berkeley, California, USA}
\date{June, 2002}
\maketitle

\parbox{12cm}{\small
We introduce a model of computation based on \emph{read only memory} (ROM),
which allows us to compare the
space-efficiency of reversible, error-free classical computation with
reversible, error-free quantum computation.
We show that a ROM-based quantum computer with one writable qubit
is universal, whilst two writable bits are required for a universal classical
ROM-based computer.
We also comment on the time-efficiency advantages of quantum computation
within this model.
}

\section{Introduction}	     

To date, the main drive of research into quantum computation has been to show
that the time requirements for solving certain problems are smaller for a
quantum computer than they are for a classical computer. Perhaps the most
well known result is Shor's algorithm\cite{Shor94}, which enables a quantum
computer to factor large integers with a subexponential speed-up over the best
known classical solution. 
Other examples of increased time-efficiency using quantum
computation are the Deutsch-Jozsa algorithm\cite{Deutsch92}, which provides an
\emph{unbounded} speed-up over classical \emph{deterministic}
algorithms\footnote{The Deutsch-Jozsa algorithm makes $O(1)$ quantum 
queries as opposed to the $\Omega(n)$ required deterministically classically.}, 
and Grover's
search algorithm\cite{Grover97a}, which provides a polynomial speed-up.
For a general introduction to quantum computation, the reader could consult
Nielsen and Chuang\cite{Nielsen00} or Preskill\cite{Preskill98}.

Whilst time is often considered the key resource to be minimized during the
solving of a problem, another resource of considerable importance is space.
Space complexity is the study of the number of (qu)bits required by a
computer to solve a problem.
At present the experimentally viable `quantum computers' have fewer than ten
qubits\cite{Sackett00,Knill00}, thus the question of what can be computed
using
small quantum computers is of interest. %needs to be addressed.
As is conventional in space complexity theory, we shall differentiate
between \emph{read-only} memory (ROM) and \emph{writable} memory\cite{Papa94}.
The space complexity will be a function of the writable memory only.
Previous work on space-bounded quantum computation has looked at quantum
Turing
machines\cite{Watrous99} and quantum finite-state automata\cite{Ambainis98},
both of which are bounded-error models.
In this paper we introduce a model which allows us to compare the space
complexity of error-free, reversible quantum and classical computation.

Some of the results discussed in this paper are related to those found in
Barenco et al. \cite{Barenco95a}, however we shall be requiring that most of
the bits in our circuits are \emph{read only}, a restriction which did not
need to be addressed in \cite{Barenco95a}. It is this restriction which allows
us to demonstrate the difference in space complexity of quantum and classical
ROM computation.

The structure of this paper is as follows.
In Section 2 we explain in detail our ROM-based computation model.
In Section 3 we show that a  ROM-based quantum
computer with one writable qubit is universal.
In Section 4 we show that two writable bits are required for
a universal classical ROM-based computer.
Finally, in Section 5 we comment on time-efficiency within the model.

\section{ROM-based Computation}\label{ROM}

In this paper we are considering mappings between strings of boolean variables
(bits) of the following form,
\be\label{map}
 u_1u_2\dots u_j\underbrace{00\dots 0}_{n \ \mathrm{(qu)bits}}
   &\xrightarrow[]{F}&
   u_1u_2\dots u_j f_1f_2\dots f_n,
\ee
where each $u_i \in \{0,1\}$ and each $f_i \in \{0,1\}$. It is evident from
\eqn{map} that the first $j$ bits have the same initial and final
values, however in our model, we shall require that the values of the first
$j$ bits are also not altered during any of the steps of the computation, so
we can consider them to be \emph{read-only memory} or ROM bits. Each of the
last $n$ bits are mapped to zero or one, depending on the values of the
ROM bits. Therefore we can think of each of these $n$ bits as
\emph{writable} bits, whose final value is a boolean function of the
ROM-bits,
\be
  f_i(u_1,u_2,\dots,u_j) &:& {\mathbb B}^j_2 \rightarrow {\mathbb B}_2 \quad i
  \in \{1,2,\dots,n\},
\ee
where ${\mathbb B}^j_2$ denotes a binary string of length $j$.
In the classical case, a given function $f_i$ is generated by a sequence of
arbitrary classical \emph{reversible}
gates acting
on the $n$ writable bits. Additionally, any of these gates can be applied
conditionally upon the value of \emph{one} of the $j$ ROM bits.
We are using only reversible gates to preserve the number of
writable bits. Any irreversible gate which increases the number of writable
bits (e.g. FANOUT) has an associated space complexity cost, whilst
irreversible gates which reduce the number of writeable bits (e.g. AND)
dissipate energy, and therefore have a thermodynamic cost \cite{Landauer61}. 

In the quantum case,
arbitrary quantum gates can be applied to the $n$ qubits, and once again any
of these gates can be applied conditionally upon the value of \emph{one} of
the $j$
ROM bits. However, it should be remembered that each of the $f_i$ are boolean
expressions, thus whilst the qubits can exist in superpositional
states during the computation, at the conclusion they must be in a
computational
basis state. This means that the entire computation (including measurement) is
deterministic and reversible, as measuring the $n$ qubits at the end of
the computation will have no effect on their state.
Intermediate
measurements can be made in neither the quantum or classical models, as the
storing of the measurement result would be effectively expanding the
workspace.

It is perhaps natural to question why we are allowing a given gate to be
conditional on only \emph{one} of the ROM bits. Generally, in both quantum
and classical computation, arbitrary numbers of controls are
allowed,
as these can always be broken down into gates containing a fixed number of
controls (two in the case of quantum computation\cite{DiVincenzo95}, and
three in the case of classical computation\cite{Fredkin82}). 
If arbitrary numbers of controls are allowed it is trivial to show
that a one (qu)bit ROM computer is universal.
However, breaking
down such conditional gates requires the conditional bits to be writable, and
therefore has an associated space complexity cost.
It should also be pointed out that there is nothing unique about allowing only
one control ROM bit per gate. The results presented in the Sections 3 and 4
would be unaffected by allowing any \emph{fixed} number of simultaneous 
conditional ROM bits. 
In particular, a one bit classical computer will still not be universal, 
as will be shown in Section 4.

\begin{figure}[ht]
\centering
 \scalebox{0.75}{\includegraphics{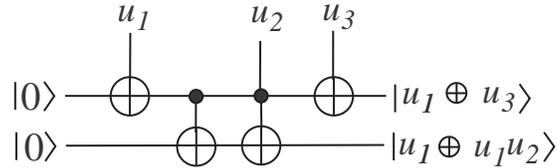}}
 \caption{An example of a ROM-based circuit diagram, the circles containing
 crosses indicate NOT gates and the black circles indicate controls. 
 The variables at the top of the diagram are the ROM bits.}
 \label{example}
\end{figure}

Throughout this paper we shall be using circuit diagrams to represent our
ROM-based computations. As is standard in quantum computational circuit
diagrams, the writable (qu)bits will be represented as horizontal lines,
whose states change as various gates are applied from left to right. The ROM
bits will be depicted above the circuit diagram, with a line from a ROM
bit to a gate implying that this gate is applied only if the ROM bit has value
one. \fig{example} contains an example of a ROM computation circuit diagram.
This diagram depicts the computation
\be\label{egeq}
 u_1u_2u_3\ket{0}\ket{0} &\xrightarrow[]{F}& u_1u_2u_3 \ket{f_1}\ket{f_2},
\ee
where
\be
  \ket{f_1(u_1,u_3)} &=& \ket{u_1 \xor u_3} \quad \mathrm{and} \nonumber \\
  \ket{f_2(u_1,u_2)} &=& \ket{u_1 \xor u_1u_2}.
\ee
Please note that we shall be using kets to denote the writable elements of a
ROM-based computer, irrespective of whether these elements are bits or qubits.

There are $2^\wedge(n2^j)$ Boolean functions from $j$ bits to $n$ bits.
We shall define as \emph{universal} a ROM-based computer which of capable of
calculating all of these functions.
In Section 3 we show that \emph{one}
writable qubit is sufficient for a universal ROM-based quantum computer,
whilst in Section 4 we show that \emph{two} writable bits are required for
a universal ROM-based classical computer.
In either the classical or quantum
case it is easy to see that if the ROM model is universal with $m$
writable (qu)bits then it is universal for any $m' \geq m$, so the main
interest is in determining the minimal $m$ for which universality holds.

The arguments contained in the following sections depend upon the fact that XOR
and conjunction produce a distinguished normal form. In order to define this
distinguished normal form, let us first review some propositional logic
theory.
It is well known that AND and negation are sufficient to
express any boolean proposition\cite{Hilbert50}. Using the three simple
equivalences,
\be\label{distribute}
  1a &\equiv& a \nonumber\\
  \bar{a} &\equiv& a \xor 1 \\
  a(b \xor c) &\equiv& ab \xor ac, \nonumber
\ee
it follows that AND and XOR are also sufficient, as every
negated sentence, $\bar{a}$, can be replaced by $a \xor 1$. This implies that
all $2^\wedge\!(2^j)$
propositions composed of $j$ boolean variables can be expressed as an XOR
disjunction of conjunctions, involving no negations. Hence, XOR and AND
produce a \emph{normal form}. XOR and AND also produce a
\emph{distinguished normal form},
as every expression involving only XOR
disjunctions of conjunctions, with no negations, is unique up to transposition
of conjunctions 
\footnote{For more information on such algebraic forms the reader should 
see for example MacWilliams and Sloane (p. 371)\cite{MacWilliams77}.}. 
To see that each
expression is unique, we note that there are exactly $\binom{j}{k}$ distinct
conjunctions involving exactly $k$ of $j$ variables. Thus, the total number of
conjunctions is $\sum_{k=0}^{j}{\binom{j}{k}} = 2^j$. The presence or absence
of each of these terms gives the $2^\wedge\!(2^j)$ different boolean
propositions.

To prove that a ROM-based computer is universal, we need to show that each
writable (qu)bit can be mapped from $0$ to any of the $2^\wedge\!(2^j)$
different boolean propositions.
As every boolean expression can be written as an XOR disjunction of
conjunctions, it is sufficient to show that we can transform
$\ket{f}$ to $\ket{f \xor u_1u_2\dots u_m}$ where $f$ is an arbitrary
boolean function and $m \in \{1,2,\dots,j\}$.

\section{One writable qubit is universal}\label{onequbit}

\begin{figure} [ht]
\centering
 \scalebox{0.50}{\includegraphics{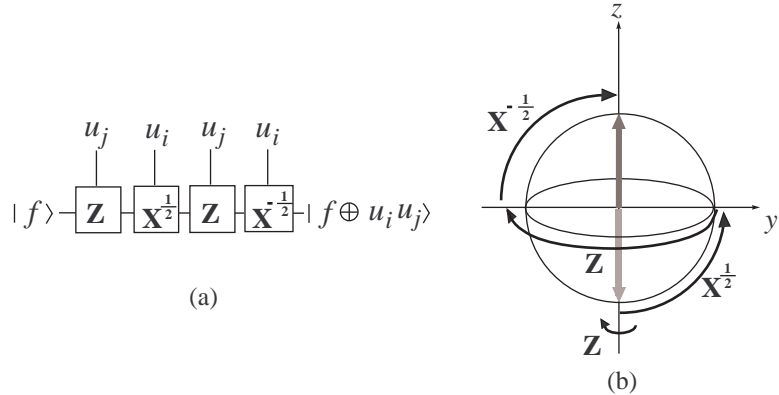}}
 \caption{(a) Circuit diagram of the ROM sequence used to transform
              $\ket{f}$ to $\ket{f \xor u_i u_j}$.
          (b) Bloch sphere representation showing the state $\ket{0}$
              transforming to the state $\ket{1}$, when $u_i = u_j = 1$.
              For all other values of $u_i$ and $u_j$, $\ket{f}$ remains
              unchanged.} \label{qtimes}
\end{figure}

We will now use the Pauli operators,
\be
 Z = \pauli{1}{0}{0}{-1} &\mathrm{and}& X = \pauli{0}{1}{1}{0},
\ee
as well as the operators
\be
 Z^{\pm\frac{1}{2}} = \pauli{1}{0}{0}{\pm i}
       &\mathrm{and}&
 X^{\pm\frac{1}{2}} = \frac{1}{2}\pauli{1 \pm i}{1 \mp i}{1 \mp i}{1 \pm i},
\ee
%$X^{-\frac{1}{2}}, X^{\frac{1}{2}}, Z^{-\frac{1}{2}}$
%and $Z^{\frac{1}{2}}$ 
to show that a ROM-based quantum computer with one
writable qubit is universal.
We denote by $W_{u_i}$ an operator $W$ which is applied conditionally
on the ROM bit $u_i$. The sequence of one-qubit gates,
\be\label{qand}
 X_{u_i}^{-\frac{1}{2}} Z_{u_j}  X_{u_i}^{\frac{1}{2}} Z_{u_j}
   &=& (i X)_{u_i,u_j}
\ee
performs a bit flip \emph{if and only if} ROM bits $u_i = u_j = 1$. 
If the ROM bits were qubits, this would be equivalent to the Toffoli gate
construction given in Section VI B of Barenco et al.\cite{Barenco95a}.
Evidently, if both $u_i$ and $u_j$ are zero, no gate is
performed, whilst if only one of $u_i$ or $u_j$ is one, then a gate is
performed, followed immediately by its inverse, leaving the writable qubit
unaltered. However, if both $u_i$ and $u_j$ are one, the sequence of four
gates
combine to give the Pauli $X$ matrix, which has the effect of flipping the
qubit in the computational basis. A circuit diagram for this computation is
depicted in \fig{qtimes}(a), whilst \fig{qtimes}(b) uses the Bloch sphere
representation\footnote{The pure state of a single qubit can always be
represented
by a point on a unit sphere, known as the Bloch sphere. For more information
see \cite{Nielsen00} page 15.} to show how a qubit initial
in the state $\ket{0}$ is transformed into the state $\ket{1}$ iff
$u_i=u_j=1$.
Thus, the sequence in \eqn{qand} takes a writable qubit from $\ket{f}$
to $\ket{f \xor u_iu_j}$.

Now each of the $Z_{u_j}$ terms in \eqn{qand} can be replaced by
\be\label{qand2}
 Z_{u_k}^{-\frac{1}{2}} X_{u_j}  Z_{u_k}^{\frac{1}{2}} X_{u_j}
   &=& i Z_{u_ku_j},
\ee
which gives the sequence
\be\label{qand3}
 X_{u_i}^{-\frac{1}{2}} Z_{u_ju_k}  X_{u_i}^{\frac{1}{2}} Z_{u_ju_k}
   &=& X_{u_iu_ju_k},
\ee
ignoring an overall phase factor. This new sequence of gates takes
$\ket{f}$ to $\ket{f\xor u_iu_ju_k}$.
By replacing the $X_{u_j}$ terms in \eqn{qand2} by sequences of
the form given in \eqn{qand} it is easy to see, by recursion, that we can
generate a sequence of gates which transforms $\ket{f}$ to $\ket{f
\xor u_1u_2\dots u_m}.$ This shows that a ROM-based quantum
computer with one writable qubit is universal.
We note that this scheme is time inefficient, it requires a number of ROM
calls which scales exponentially with the number of ROM bits, however in
Section 5 we will introduce a time efficient scheme.

\section{Two writable bits are Universal}\label{twobit}

A ROM-based classical computer with one writeable bit, and the ability to
apply a NOT gate conditioned on any fixed number of ROM bits will not be
universal. This can be seen as a consequence of Theorem 5.2 from
\cite{Toffoli80}, which states that there exist invertible functions of order
$n$ which cannot be obtained by composition of generalized Toffoli gates of
order strictly less than $n$.

\begin{figure} [ht]
\centering
 \scalebox{0.85}{\includegraphics{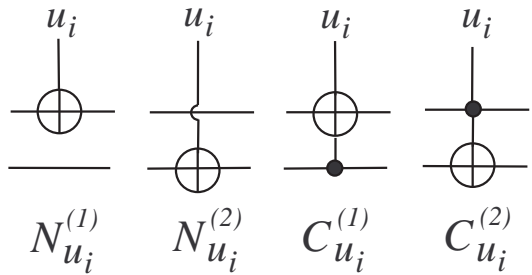}}
 \caption{Circuit diagram representation of the four transforms given in
          \eqn{ctransforms}.}
 \label{crom}
\end{figure}

Now consider a ROM-based classical computer with two writable bits. It is
possible to deduce that this will be universal using Lemma 7.3 from Barenco et
al. \cite{Barenco95a}. Here, we show that a two-bit ROM-based classical
computer is universal using
the four gates depicted in \fig{crom}, which perform the transforms
\begin{subequations}
  \label{ctransforms}
\be
 \ket{\alpha}\ket{\beta} &\xrightarrow[]{N^{(1)}_{u_i}}&
              \ket{\alpha\xor u_i}\ket{\beta}  \\
 \ket{\alpha}\ket{\beta} &\xrightarrow[]{N^{(2)}_{u_i}}&
              \ket{\alpha}\ket{\beta\xor u_i}  \\
 \ket{\alpha}\ket{\beta} &\xrightarrow[]{C^{(1)}_{u_i}}&
              \ket{\alpha\xor u_i \beta}\ket{\beta}  \\
 \ket{\alpha}\ket{\beta} &\xrightarrow[]{C^{(2)}_{u_i}}&
              \ket{\alpha}\ket{\beta\xor u_i \alpha}.
\ee
\end{subequations}

We now wish to show, using the four transforms from \eqn{ctransforms}
that it is possible to transform the writable bits from the state
$\ket{\alpha}\ket{\beta}$ to $\ket{\alpha}\ket{\beta\xor u_1u_2\dots u_m}$.
Let us denote by $S_0$ the gate $N^{(1)}_{u_1}$, which takes
$\ket{\alpha}\ket{\beta}$ to $\ket{\alpha\xor u_1}\ket{\beta}$.
It is not hard to show that the sequence
\be
 S_1 &:& C^{(2)}_{u_2} S_0 C^{(2)}_{u_2} S_0
\ee
performs the transform
\be
 \ket{\alpha}\ket{\beta} \xrightarrow[]{S_1} \ket{\alpha}\ket{\beta\xor
 u_1u_2}.
\ee
Now, suppose we have a sequence of gates, $S_{m-1}$, which performs the
transform
\be
 \ket{\alpha}\ket{\beta} &\xrightarrow[]{S_{m-1}}&
    \ket{\alpha}\ket{\beta \xor u_1 u_2 \dots u_{m-1}}.
\ee
Then there exists a sequence of gates,
\be
 S_m &:& C^{(1)}_{u_m} S_{m-1} C^{(1)}_{u_m} S_{m-1}
\ee
which perform the transform
\be
 \ket{\alpha}\ket{\beta} &\xrightarrow[]{S_m}&
    \ket{\alpha \xor u_1 u_2 \dots u_m}\ket{\beta}.
\ee
Therefore a ROM-based classical computer with two writable bits is universal.

\section{Time efficiency}\label{time}

%If we assume that we can only apply quantum gates with some fixed, finite
%precision, then 
A simple counting argument shows that there exists boolean
expressions which will require an exponential number of ROM calls on either a
quantum or classical ROM computer with a fixed number of writeable (qu)bits.
However, it is an open question as to whether there exist specific
boolean expressions which can be generated on a one qubit quantum computer
using a polynomial number of ROM calls, which require an exponential number
of ROM calls on a two bit classical computer.
Here, we look at one function in particular, and show that it can be computed
efficiently on a one qubit quantum ROM computer.
Whether this function can be computed efficiently on a two bit classical ROM
computer rests on an unanswered question in classical complexity theory.

Consider the transform
\be
 \ket{f} &\xrightarrow[]{F}& \ket{f \xor u_1u_2\dots u_j}.
\ee
\eqn{qand} indicates that the transform
$\ket{f} \rightarrow \ket{f \xor u_1u_2}$
can be accomplished using four ROM calls. Now, by making the following
replacements,
\begin{subequations}
\be
 X^{-\frac{1}{2}}_{u_1}  &\mathrm{with}&
 X_{u_1}^{-\frac{1}{4}} Z_{u_2}  X_{u_1}^{\frac{1}{4}} Z_{u_2} \\
 X^{\frac{1}{2}}_{u_1}  &\mathrm{with}&
 X_{u_1}^{\frac{1}{4}} Z_{u_2}  X_{u_1}^{-\frac{1}{4}} Z_{u_2}  \\
 Z_{u_2}  &\mathrm{with}&
 Z_{u_3}^{-\frac{1}{2}} X_{u_4}  Z_{u_3}^{\frac{1}{2}} X_{u_4},
\ee
\end{subequations}
we can transform $\ket{f} \rightarrow \ket{f \xor u_1u_2u_3u_4}$ using 16 ROM
calls.
A direct extension of this method, replacing each $X^{\pm1/2^n}$ by
\be
 X^{\pm1/2^{n+1}} Z  X^{\mp1/2^{n+1}} Z,
\ee
and each $Z^{\pm1/2^n}$ by
\be
 Z^{\mp1/2^{n+1}} X Z^{\pm1/2^{n+1}} X,
\ee
allows us to take the AND of up to $2^k$ ROM
bits using exactly $4^k$ ROM calls. Thus, to take the AND of $O(j)$ ROM bits
requires only $O(j^2)$ quantum gates. (Note that if the number of ROM bits is
not a power of two we need simply include some dummy ROM bits set equal to 1.)

Now let us consider the classical case. First, suppose that we have a three
bit classical ROM computer. We wish to efficiently perform the transform
\be
 \ket{f}\ket{g}\ket{h} &\xrightarrow[]{F}& \ket{f \xor u_1u_2\dots u_j}\ket{g}\ket{h}.
\ee
Another way of stating this transform, is that we wish to perform the 
permutation,
\be\label{perm}
 (0 \ 1)(2 \ 3)(4 \ 5)(6 \ 7)
\ee
on the eight states of the writable bits,
if all the ROM bits are in the state $1$, and perform the identity permutation
otherwise.
The permutation in \eqn{perm} can be generated by applying the following five
state permutations in succession from left to right,
\be\label{5perm}
 (0\ 1\ 2\ 5\ 4) \ (0\ 4\ 5\ 3\ 2) \ (4\ 5\ 6\ 1\ 0) \ (4\ 0\ 1\ 7\ 6)
\ee
where each of these permutations is applied if and only if all ROM bits are in 
the $1$ state. Now suppose that $\rho$ is a permutation of $5$ states, and $F$ 
is a Boolean function that can be computed by a depth $d$ Boolean circuit 
consisting of NOT gates and 2-input, 1-output AND/OR gates. Then Barrington's
Theorem \cite{Barrington89} states that there is a permutation branching
program\footnote{A permutation branching program is the same as the classical ROM model of 
computation
except that the workspace can be in one of $k$ states, and each step of the 
computation is a permutation of the states conditional on the ROM bit.}
of length $\leq 4^d$ on 5 states such that:
\begin{itemize}
\item it maps every state to itself if $F=0$.
\item it permutes the states according to $\rho$ of $F=1$.
\end{itemize}
As the AND of $j$ Boolean variables can be computed by a Boolean circuit of 
depth $\lceil \log_2 j \rceil$, this theorem, and \eqn{5perm} indicates that a 
three bit ROM classical computer can perform $F$ using $O(j^2)$ ROM calls. 

A two bit classical ROM computer is equivalent to a 4 state permutation
branching program. The power of a 4 state permutation program is 
unknown\cite{Barrington89}.
However, here we conjecture that the transform
\be
 \ket{0}\ket{0} &\xrightarrow[]{F^\prime}& \ket{u_1u_2\dots u_j}\ket{0},
\ee
requires a number of ROM calls which scales exponentially with $j$.
There is sequence of gates which can perform $F^\prime$ on a two bit 
classical ROM computer using exactly $R(j)$ ROM calls, where
\be\label{recurs}
 R(j) &=& R(j\! -\! 1) + 2^{\lfloor j/2 \rfloor}, \quad R(1) = 1.
\ee 
This is clearly exponential in $j$, and exhaustive numerical searches for 
$j < 5$ have shown \eqn{recurs} to be minimal.

\section{Discussion}

In conclusion, we have introduced a model, which allows the
comparison of space-efficiency between error-free, reversible quantum and
classical computation. We have shown that quantum computation is more space
efficient within this model, requiring only one qubit for universality, as
opposed to two bits.
We have also conjectured that the minimal quantum ROM computer can calculate
certain boolean functions exponentially faster than the minimal classical ROM
computer.

It would be interesting to compare the classical and quantum models, allowing
for bounded-error computation, that is, the writeable bits are mapped to
the correct boolean functions of the ROM bits with some probability
$1-\epsilon$. Preliminary investigations indicate  that the quantum model
would
still be more powerful than the classical model.

\end{document}